\documentclass[10pt]{article}

\usepackage{amsmath}
\usepackage{amssymb}

\usepackage{graphicx}

\usepackage{cite}

\usepackage{color}

\usepackage{setspace}
\usepackage[labelfont=bf,labelsep=period,justification=raggedright]{caption}

\makeatletter
\renewcommand{\@biblabel}[1]{\quad#1.}
\makeatother

\date{}

\usepackage{lineno}
\linenumbers
\usepackage{url}  

\begin{document}

\begin{flushleft}
{\Large
\textbf{Alignment-free comparison of next-generation sequencing data using compression-based distance measures}
}
\\
Ngoc Hieu Tran$^{\ast}$,
Xin Chen$^{}$
\\
Division of Mathematical Sciences, School of Physical and Mathematical Sciences, Nanyang Technological University, Singapore
\\
$\ast$ E-mail: nhtran@ntu.edu.sg
\end{flushleft}

\section*{Abstract}

Enormous volumes of short reads data from next-generation sequencing (NGS) technologies have posed new challenges to the area of genomic sequence comparison.
The multiple sequence alignment approach is hardly applicable to NGS data due to the challenging problem of short read assembly.
Thus alignment-free methods need to be developed for the comparison of NGS samples of short reads.
Recently, new $k$-mer based distance measures such as {\it CVTree}, $d_{2}^{S}$, {\it co-phylog} have been proposed to address this problem.
However, those distances depend considerably on the parameter $k$, and how to choose the optimal $k$ is not trivial since it may depend on different aspects of the sequence data.
Hence, in this paper we consider an alternative parameter-free approach: compression-based distance measures.
These measures have shown impressive performance on long genome sequences in previous studies, but they have not been tested on NGS short reads.
In this study we perform extensive validation and show that the compression-based distances are highly consistent with those distances obtained from the $k$-mer based methods, from the alignment-based approach, and from existing benchmarks in the literature.
Moreover, as these measures are parameter-free, no optimization is required and they still perform consistently well on multiple types of sequence data, for different kinds of species and taxonomy levels.
The compression-based distance measures are assembly-free, alignment-free, parameter-free, and thus represent useful tools for the comparison of long genome sequences and NGS samples of short reads.


\section*{Introduction}

%
Recent advances in next-generation sequencing (NGS) technologies have produced massive amounts of short reads data, bringing up new promising opportunities in many research areas of biomedical sciences such as RNA-seq, ChIP-seq, {\it de novo} whole genome sequencing, metagenome shotgun sequencing, etc \cite{metzker_09}.
This new type of data also poses interesting challenges to the field of genomic sequence analysis, including the problem of sequence comparison.
Sequence distance measures are often applied to compare 16S rRNA sequences, mtDNA sequences, individual or multiple genes, or even whole genome sequences \cite{waterman_95,durbin_99,vinga_03}.
The obtained distances then can be used for the clustering and classification problems, for the reconstruction of phylogenetic trees, for the studies of the evolution and relationship of species, etc.
Distance measures have also been developed for the comparison and classification of metagenomic samples in the studies of microbial communities \cite{kuczynski_10}.
However, with the development of NGS technologies, the new type of sequence data emerges: NGS short reads are orders of magnitudes shorter than long genome sequences (that is, 16S rRNA sequences, mtDNA sequences, whole genome sequences), and more importantly, they can be generated at unprecedented high throughput.
Hence, it is highly desirable to go beyond the comparison of long genome sequences to develop new methods for the comparison of NGS samples of millions of short reads \cite{chan_13}.
%

%
Sequence comparison methods can be classified into two categories: alignment-based and alignment-free.
An alignment-based method first performs a multiple sequence alignment (MSA) of the input sequences and then computes their pairwise distances using nucleotide substitution models \cite{waterman_95,durbin_99}.
%
%
Due to computational limitations, this MSA approach, however, is often only applicable to 16S rRNA sequences, mtDNA sequences, individual or a few genes, and it hardly can be applied to whole genome sequences.
The problem is even more challenging for NGS short reads as one needs to assemble the short reads to create full-length sequences before performing multiple alignment of the assembled sequences.
Short read assembly is well-known as a challenging problem, especially for species without reference genomes ({\it de novo} assembly).
Short reads produced from metagenomic samples are even more difficult to handle as each sample is a mixture of different genomes.
Finally, even if an assembly can be done, assembled sequences are likely to be error-prone and hence their multiple alignment may be not reliable for further comparison purpose.

%
Alignment-free methods have been proposed to overcome the limitations of the alignment-based approach \cite{vinga_03}.
Their key advantage is the scalability to whole genome sequences, even those large genomes up to gigas base pairs.
A wide class of alignment-free methods uses the observed $k$-mers ($k$-tuples, $k$-words) of the input sequences to measure their pairwise distances \cite{qi_04,xu_09,reinert_09,wan_10,yi_13}.
Markov models were also proposed for DNA sequence comparison \cite{pham_04}, and they can be incorporated with $k$-mer distributions to provide more accurate distances \cite{kantorovitz_07,dai_08}.
Another class of alignment-free distances arises from information theory, in particular, from data compression.
Those distances are calculated from the relative information between the input sequences using Kolmogorov complexity \cite{li_01,li_04,keogh_04}, Lempel-Ziv complexity \cite{otu_03}.
Unlike $k$-mer based distance measures which depend on the parameter $k$, compression-based distance measures are parameter-free and hence more consistent.
They have been successfully applied to many clustering and classification problems with several types of data, including DNA sequences, texts and languages, time series, images, sound, video \cite{li_01,li_04,benedetto_02,otu_03,keogh_04,ito_10}.

%
Interestingly, recent studies have suggested that alignment-free methods can also be applied directly to NGS short reads without prior assembly.
In particular, the following three $k$-mer based measures have shown impressive performance on both NGS short reads and long genome sequences: {\it CVTree} \cite{qi_04,xu_09}, $d_{2}^{S}$ \cite{jiang_12,song_13} and {\it co-phylog} \cite{yi_13}.
For a given $k$, {\it CVTree} and $d_{2}^{S}$ measure the distance between two DNA sequences (or two NGS samples) based on the observed frequencies of $k$-mers.
{\it co-phylog}, on the other hand, computes the distance from the average nucleotide substitution rate in the observed $k$-mers.
However, these measures may depend considerably on the parameter $k$, and wrong choice of $k$ may lead to inconsistent results.
In principle, larger values of $k$ allow the measures to use more parameters in their background models to better capture the characteristics of the sequences (or samples).
However, not sufficient information from the observed data will lead to poor estimation of those parameters and result in inaccurate distances.
Hence, the optimal $k$ may depend on the type of sequence data, the species of interest, or the taxonomy level.
When the measures are applied to NGS short reads, even more factors need to be considered such as the NGS platform, the sequencing depth, the read length, etc.
Thus, how to choose the optimal $k$ is not trivial.
In the worst case, one may need to try all possible values of $k$ to find the best distances according to some given benchmark. 
However, in many real applications when there is no benchmark available, it is difficult to identify the best result and the corresponding optimal $k$.

Hence, in this study we want to extend the compression-based approach to NGS short reads.
The advantages of this approach are assembly-free, alignment-free, and more importantly, parameter-free.
We demonstrate that the compression-based distance measures developed in \cite{li_01,li_04,keogh_04} can be successfully applied to both NGS short reads and long genome sequences (including 16S rRNA, mtDNA, and whole genome sequences).
Extensive validation was conducted to assess the accuracy of the sequence distance measures, compression-based and $k$-mer based, on four data sets: 29 mammalian mtDNA sequences, 29 {\it Escherichia/Shigella} genomes, 70 {\it Gammaproteobacteria} genomes, and 39 mammalian gut metagenomic samples.
The data sets include various types of genomic sequences, in silico and real NGS short reads, different species and taxonomy levels.
The validation results show that the compression-based distances are highly consistent with those distances obtained from the $k$-mer based methods, from the MSA approach, and from existing benchmarks in the literature.
The results also show that the $k$-mer based distance measures depend remarkably on the choice of $k$, and the optimal $k$ varies across different data sets.
The compression-based distance measures, on the other hand, are parameter-free and are applicable to all data sets without any modification or optimization.
Moreover, their results obtained from different types of sequences and short reads of the same species are highly consistent with each other.
To the best of our knowledge, our work is the first study to assess the accuracy of the compression-based distance measures on NGS short reads, demonstrating their accuracy and consistency.
An implementation of the compression-based distance measures is available at \url{http://www1.spms.ntu.edu.sg/~chenxin/GenCompress/}.
Detailed results are presented in the following sections.

\section*{Materials and Methods}

\subsection*{Compression-based distance measures}

We used the compression-based distance measures that were proposed in \cite{li_01,li_04,keogh_04}.
Those measures were first developed based on the theory of Kolmogorov complexity \cite{li_08}.
%
%
As Kolmogorov complexity is not computable, the authors further refined the measures using data compression.
In particular, the following distance measure was proposed in \cite{li_01}:

\begin{equation}\label{eqn:distance_d}
  d(x,y) = \frac{C(x|y)+C(y|x)}{C(xy)}. 
\end{equation}

Here $C(x)$ denotes the size of the compressed file of sequence $x$ from a standard compressor,
$xy$ denotes the concatenation of two sequences $x$ and $y$,
and $C(x|y)$ denotes the size of the compressed file of sequence $x$ conditioning on sequence $y$.
The authors further proposed a more mathematically precise version in \cite{li_04} which was referred to as normalized compression-based distance (NCD):

\begin{equation}\label{eqn:distance_NCD}
  d^{\scriptsize \mbox{NCD}}(x,y) = \frac{\mbox{max}\{C(x|y),C(y|x)\}}{\mbox{max}\{C(x),C(y)\}}. 
\end{equation}

Keogh {\it et al.} (2004) also proposed a simplified version of $d$ called compression-based dissimilarity measure (CDM):

\begin{equation}\label{eqn:distance_CDM}
  d^{\scriptsize \mbox{CDM}}(x,y) = \frac{C(xy)}{C(x)+C(y)}. 
\end{equation}

A good compressor should be able to remove redundant information that is shared between two sequences $x$ and $y$ when compressing their concatenation or compressing one sequence by conditioning on the other.
Hence, if two sequences $x$ and $y$ are identical, we have $C(x|y) \simeq 0$, $C(y|x) \simeq 0$, and $C(xy) \simeq C(x)=C(y)$.
Then $d^{\scriptsize \mbox{NCD}}$ and $d^{}$ are approximately equal to zero, while $d^{\scriptsize \mbox{CDM}}$ is about $\frac{1}{2}$.
On the other hand, if two sequences $x$ and $y$ share no information, we have $C(x|y) \simeq C(x)$, $C(y|x) \simeq C(y)$, $C(xy) \simeq C(x)+C(y)$, and hence all three distances are approximately equal to one.
More detailed analysis and properties of the compression-based distance measures and Kolmogorov complexity can be found in \cite{li_04}.

%
We used the tool {\it GenCompress} \cite{chen_99} for compression.
The advantage of {\it GenCompress} is that it can perform conditional compression $x|y$.
%
%
When applying {\it GenCompress} to an NGS sample, we simply concatenated all short reads of the sample to form a single sequence and then compressed that sequence.
Using the same compression tool allows a consistent and fair comparison across different data sets.

\subsection*{$k$-mer based distance measures}

We considered three $k$-mer based distance measures: {\it CVTree} \cite{qi_04,xu_09}, $d_{2}^{S}$ \cite{jiang_12,song_13} and {\it co-phylog} \cite{yi_13}.
Given a fixed length $k$ and two DNA sequences (or two NGS samples), {\it CVTree} measures the correlation distance between their composition vectors, where each composition vector is the collection of the normalized frequencies of $k$-mers in the corresponding sequence (or sample).
The $d_{2}^{S}$ distance is an NGS-extension of the $D_{2}$, $D_{2}^{*}$, and $D_{2}^{S}$ statistics which were proposed in \cite{reinert_09,wan_10} for the comparison of long genome sequences.
%
%
The main difference between $d_{2}^{S}$ and {\it CVTree} lies in the normalization of the frequency vectors of $k$-mers.
%
%
The {\it co-phylog} distance measure is also based on $k$-mers but not in the frequency context.
It first constructs context-object structures from the observed $k$-mers and then measures the distance as the number of structures that have the same contexts, but different objects in the two sequences (or two samples).
In other words, this is a micro-alignment process that attempts to estimate the average nucleotide substitution rate between two DNA sequences (or two NGS samples).

%
The implementations of {\it CVTree} and $d_{2}^{S}$ have the option to input the parameter $k$.
Hence, we were able to try all possible values of $k$ allowed by the programs.
For {\it CVTree}, we tried $k$ from 2 to 32.
However, for $d_{2}^{S}$, we were not able to run the program for $k>9$ due to some ``segmentation fault''.
There is no input option for {\it co-phylog}, thus we simply used its default settings.

\subsection*{Accuracy assessment}

We examined the above six alignment-free distance measures $d^{\scriptsize \mbox{NCD}}$, $d^{}$, $d^{\scriptsize \mbox{CDM}}$, {\it CVTree}, $d_{2}^{S}$, and {\it co-phylog} on both NGS short reads and long genome sequences (including 16S rRNA, mtDNA, and whole genome sequences).
The sequences and short reads were retrieved and simulated from four data sets: 29 mammalian mtDNA sequences \cite{li_04,dai_08,otu_03,cao_98}, 29 {\it Escherichia/Shigella} genomes \cite{yi_13}, 70 {\it Gammaproteobacteria} genomes \cite{yi_13}, and 39 metagenomic mammalian gut samples \cite{muegge_11,jiang_12}.

The tool {\it MetaSim} \cite{richter_08} was used to simulate short reads from genome sequences.
It offers four error models: 454, Sanger, Empirical, and Exact, which correspond to different NGS platforms and the non-error case.
We set the read length to be 100 and used default settings for other parameters.
Short reads were simulated at four sampling depths: $1\times$, $5\times$, $10\times$, and $30\times$.

To evaluate the accuracy of the alignment-free distances, we compared them with those obtained from the MSA approach if applicable.
The tool Clustal Omega \cite{sievers_11} was used to perform MSA of mtDNA and 16S rRNA sequences, and then the tool {\it dnadist} in the package PHYLIP \cite{PHYLIP} was used to calculate the distance matrix from the MSA.
Moreover, we also compared the alignment-free distances with available benchmarks in existing literatures.

We computed the correlation between each alignment-free distance and the MSA/benchmark distance to evaluate their consistency.
We also assessed the alignment-free distances by examining their corresponding phylogenetic trees.
For each distance matrix, the tool {\it neighbor} in the package PHYLIP was used to construct the phylogenetic tree using the neighbor joining method \cite{saitou_87}.
Subsequently, the tool {\it treedist} in the package PHYLIP was used to calculate the symmetric difference between the resulting tree from each alignment-free distance and the corresponding MSA/benchmark tree.
The tool TreeGraph 2 \cite{stover_10} was used to plot the trees.

Finally, to assess the clustering and classification ability of the alignment-free distances for a large number of genomes and for the case of metagenomic samples, we used the parsimony score to measure how different a clustering tree is from the true classification  (tools {\it TreeClimber} \cite{schloss_06}, {\it mothur} \cite{schloss_09}).
%
%
If a clustering tree is perfect, its parsimony score is equal to the number of groups of the true classification minus one.
The higher the parsimony score is, the more different the clustering tree is from the true classification.
The parsimony score of a clustering tree was also compared with that of randomly joined trees to calculate $p$-value and assess its statistical significance.

\section*{Results and Discussion}


\subsection*{Phylogenetic trees reconstructed from mammalian mtDNA sequences and their NGS short reads reconfirm the hypothesis (Rodents, (Ferungulates, Primates))}

The key advantage of the alignment-free distance measures over the alignment-based approach is their scalability to large data sets of whole genome sequences and NGS short reads.
However, in this section we first want to test their accuracy on a small, but very well-studied data set of 29 mammalian mtDNA sequences.
This data set has been widely used for validation in existing literatures \cite{li_04,dai_08,otu_03,cao_98} and hence reliable benchmarks are available.

\subsubsection*{Performance on mtDNA sequences}

First, we applied the six alignment-free distance measures $d^{\scriptsize \mbox{NCD}}$, $d^{}$, $d^{\scriptsize \mbox{CDM}}$, {\it CVTree}, $d_{2}^{S}$, and {\it co-phylog} to the mtDNA sequences and compared their results with those obtained from the MSA method.
Figure S1 and Table S1 show that the compression-based distances $d^{\scriptsize \mbox{NCD}}$, $d^{}$, $d^{\scriptsize \mbox{CDM}}$ and the $k$-mer based distances {\it CVTree}, $d_{2}^{S}$ (for optimal choices of $k$) show good agreement with the MSA distance, in terms of both tree symmetric difference and distance correlation.
Moreover, the phylogenetic trees reconstructed from those distances are highly consistent with existing benchmarks in the literature \cite{li_04,dai_08,otu_03,cao_98}.
The {\it co-phylog} measure, however, failed for this data set.
One possible explanation is that {\it co-phylog} may be only suitable for closely related species, as the authors have mentioned in \cite{yi_13}.
We also noted that the {\it CVTree} and $d_{2}^{S}$ distances varied considerably with respect to $k$ (Table S2).
For instance, the optimal symmetric difference between the {\it CVTree} trees and the MSA tree was 6 (for $k=10$), but the worst case was up to 48 (for $k=16,17$).
Similarly, the optimal correlation between the $d_{2}^{S}$ distances and the MSA distance was 0.88 (for $k=8$), but the lowest case was down to 0.41 (for $k=3$).

\subsubsection*{Performance on NGS short reads}

Next, we asked if similar results could be obtained from the comparison of NGS samples of short reads.
We used the program {\it MetaSim} \cite{richter_08} to simulate short reads from the mtDNA sequences with four error models 454, Exact, Empirical, Sanger, and four different sampling depths $1\times$, $5\times$, $10\times$, $30\times$.
The read length was set at 100 bp.
We used $k=10$ for the {\it CVTree} distance and $k=8$ for the $d_{2}^{S}$ distance, as suggested by their optimal performance on the mtDNA sequences in the previous section.
Since the MSA method is not applicable to NGS short reads, we still kept the MSA distance and tree obtained from the mtDNA sequences as benchmark.
At the $1\times$ sampling depth, we found that the alignment-free results were considerably different from the MSA benchmark due to the low coverage.
However, at the $5\times$ sampling depth, all five measures $d^{\scriptsize \mbox{NCD}}$, $d^{}$, $d^{\scriptsize \mbox{CDM}}$, {\it CVTree}, and $d_{2}^{S}$ produced comparably accurate results as when they were applied to the mtDNA sequences.
Further increasing the sampling depth to $10\times$, $30\times$ did not significantly improve the accuracy of the distances.

Table 1 summarizes the results for the $5\times$ sampling depth, similar results for $1\times$, $10\times$, and $30\times$ can be found in Table S2.
%
%
As shown in Table 1, the $d_{2}^{S}$ distance achieved the highest correlation with the MSA distance, followed by the {\it CVTree} and the compression-based distances.
%
%
In terms of the symmetric difference from the MSA tree, the $d^{\scriptsize \mbox{CDM}}$ distance performed consistently well for all four error models, followed by the $d_{2}^{S}$ and $d^{\scriptsize \mbox{NCD}}$ distances.
Figure 1 shows an example of the phylogenetic trees reconstructed from the $d^{\scriptsize \mbox{CDM}}$, {\it CVTree}, and $d_{2}^{S}$ distances for the NGS short reads simulated using the Empirical error model.
The $d^{\scriptsize \mbox{CDM}}$ and {\it CVTree} trees are almost identical to the MSA tree (Figure S1a) and existing benchmarks in the literature \cite{li_04,dai_08,otu_03,cao_98}, supporting the hypothesis (Rodents, (Ferungulates, Primates)).
%
%
The $d_{2}^{S}$ tree, however, has more inconsistent branches in the group Ferungulates, although it has the highest correlation with the MSA distance.
%
%
Last but not least, we also found that the alignment-free results obtained from the simulated NGS short reads were highly consistent with their corresponding counterparts obtained from the mtDNA sequences in the previous section, especially for the Exact model (Table 2).

In general, our analysis has shown that all five alignment-free distance measures $d^{\scriptsize \mbox{NCD}}$, $d^{}$, $d^{\scriptsize \mbox{CDM}}$, {\it CVTree}, and $d_{2}^{S}$ can be successfully applied to both mtDNA sequences and their NGS short reads.
The distances obtained from the NGS short reads were highly consistent with their counterparts obtained from the mtDNA sequences, and they all had good agreement with the MSA distance as well as with existing benchmarks in the literature.
The compression-based measures $d^{\scriptsize \mbox{NCD}}$, $d^{}$ and $d^{\scriptsize \mbox{CDM}}$ produced comparably accurate distances as those optimal results obtained from the $k$-mer based measures {\it CVTree} and $d_{2}^{S}$.
The {\it CVTree} and $d_{2}^{S}$ distances varied considerably with respect to $k$.
The optimal $k$ was selected according to the MSA benchmark.
This may pose a challenging problem when there is no existing benchmark available for validation.
In contrast, the compression-based measures need no optimization and performed consistently well on all data sets.

\subsection*{Phylogeny of closely related {\it Escherichia/Shigella} genomes}

In this section we assess the accuracy of the alignment-free distance measures on a data set of 29 {\it Escherichia/Shigella} genomes.
Two fundamental differences between this data set and the previous one are: (i) it consists of whole genome sequences and (ii) the species are very closely related bacteria in the genus {\it Escherichia} and the genus {\it Shigella}.
This data set has been studied previously in \cite{yi_13,zhou_10} and the authors have shown that the {\it co-phylog} distance was highly consistent with the MSA distance in terms of both tree symmetric difference and distance correlation.
Hence, to avoid the time-consuming MSA, we used the {\it co-phylog} distance as benchmark.

\subsubsection*{Performance on whole genome sequences}

We first applied the five measures $d^{\scriptsize \mbox{NCD}}$, $d^{}$, $d^{\scriptsize \mbox{CDM}}$, {\it CVTree}, and $d_{2}^{S}$ to the whole genome sequences and compared their results with the benchmark obtained from the {\it co-phylog} measure.
Table 3 shows that the $d^{\scriptsize \mbox{CDM}}$ distance performed the best in terms of both tree symmetric difference and distance correlation.
The results of $d^{\scriptsize \mbox{NCD}}$ and $d^{}$ were also considerably better than the optimal results of {\it CVTree} and $d_{2}^{S}$, especially with the remarkably high distance correlation.
The $d_{2}^{S}$ distance failed for this data set and its correlation with the benchmark {\it co-phylog} distance was significantly lower than that of the other measures.
The {\it CVTree} distance achieved good correlation but produced inconsistent phylogenetic trees for different values of $k$.
The most significant inconsistency among them is whether the genus {\it Shigella} violates the monophyleticity of the genus {\it Escherichia} or the monophyleticity of {\it E.coli} strains (Figure S2).
This was also mentioned previously in \cite{yi_13}.
%


\subsubsection*{Performance on NGS short reads}

Next, we tested the measures on the data sets of NGS short reads which were simulated from the whole genome sequences.
We used {\it MetaSim} with four error models and different sampling depths as described earlier.
Interestingly, even at the very low $1\times$ sampling depth we already obtained accurate results from the three compression-based distances $d^{\scriptsize \mbox{NCD}}$, $d^{}$, and $d^{\scriptsize \mbox{CDM}}$.
Figure 2 shows the case of the $d^{\scriptsize \mbox{CDM}}$ distance in which the phylogenetic tree reconstructed from the NGS short reads (Exact error model) is almost identical to the tree reconstructed from the whole genome sequences.
Moreover, both trees are highly consistent with the benchmark {\it co-phylog} tree.
The main difference is that in the benchmark {\it co-phylog} tree the group of {\it S.boydii} and {\it S.sonnei} was clustered with {\it E.coli} first, whereas in the $d^{\scriptsize \mbox{CDM}}$ trees that group was clustered with {\it S.flexneri} first.
Table 4 clearly shows that the compression-based distances $d^{\scriptsize \mbox{NCD}}$, $d^{}$, and $d^{\scriptsize \mbox{CDM}}$ considerably outperformed the {\it CVTree} and $d_{2}^{S}$ distances for all NGS data sets, in terms of both tree symmetric difference and distance correlation.
%
%
We also noted that while the results of the compression-based distances for the whole genome sequences (Table 3) and for the NGS short reads (Table 4) were comparable, the performance of the {\it CVTree} and $d_{2}^{S}$ distances became worse when they were applied to the NGS short reads.
Similar results for the NGS data sets obtained from the $5\times$ sampling depth can be found in Table S3.

In summary, the compression-based distances $d^{\scriptsize \mbox{NCD}}$, $d^{}$, and $d^{\scriptsize \mbox{CDM}}$ achieved comparable accuracy as the benchmark {\it co-phylog} distance on both types of data, whole genome sequences and NGS short reads, of 29 {\it Escherichia/Shigella} bacteria.
They outperformed the other two $k$-mer based distances {\it CVTree} and $d_{2}^{S}$, which either failed or produced inconsistent results for different values of $k$.
The results in this section further emphasize the wide applicability and the consistency of the compression-based distances.
They have shown to be useful measures for accurate comparison of different types of genome sequences and NGS short reads data, for both mammalian and bacteria species.

\subsection*{Classification of 70 genomes in the class {\it Gammaproteobacteria} into their correct orders}

The last section has focused on closely related bacteria at the genus level.
We next applied the MSA and the alignment-free distance measures to a larger and more complicated data set at a higher taxonomy level.
In particular, the data set consists of 70 genomes that were randomly chosen from 15 orders of the class {\it Gammaproteobacteria} (Table S4).
As the number of genomes is large and they come from different groups, it is interesting to ask if the distance measures can cluster and classify those genomes into their correct orders.
We used the parsimony score to measure the difference between a clustering tree and the true classification \cite{schloss_06}.
%
%
As the number of groups is 15, the optimal parsimony score is 14.
The higher the parsimony score is, the more different the clustering tree is from the true classification.
The parsimony score of a clustering tree was also compared with that of randomly joined trees to assess its $p$-value and statistical significance.
For this data set a parsimony score lower than 40 corresponds to a $p$-value less than 0.001 (10,000 random trees were generated).
This data set has been studied previously in \cite{yi_13} and the authors found that the {\it co-phylog} distance did not perform well because the bacteria of interest are not closely related.

\subsubsection*{Performance on 16S rRNA sequences and whole genome sequences}

As it is challenging to perform MSA of 70 whole genome sequences, we applied the MSA method to 16S rRNA sequences of those 70 genomes to obtain the benchmark distance and clustering tree (Figure S3, parsimony score = 18).
%
%
We then applied all six alignment-free distance measures $d^{\scriptsize \mbox{NCD}}$, $d^{}$, $d^{\scriptsize \mbox{CDM}}$, {\it CVTree}, $d_{2}^{S}$, and {\it co-phylog} to the 16S rRNA sequences.
Table 5 shows that the alignment-free distances were highly correlated with the MSA distance and they all achieved similar parsimony scores (17-18), except for the {\it co-phylog} distance.
Overall, the $d^{\scriptsize \mbox{NCD}}$ distance performed the best in terms of parsimony score, tree symmetric difference, and distance correlation.
Its clustering tree (Figure 3) shows that the genomes of the six orders {\it Aeromonadales}, {\it Enterobacteriales}, {\it Legionellales}, {\it Pasteurellales}, {\it Vibrionales}, and {\it Xanthomonadales} were all correctly classified into their groups.
Majority of the genomes in the remaining orders were also well clustered.
%
%
Then, we applied the alignment-free distance measures to the whole genome sequences, and the results were slightly worse than those obtained from the 16S rRNA sequences (Table 5).
We also noted that the optimal $k$ of {\it CVTree} and $d_{2}^{S}$ for the whole genome sequences were different from those for the 16S rRNA sequences (Table S5).

\subsubsection*{Performance on NGS short reads}

Finally, we applied all six alignment-free distance measures  $d^{\scriptsize \mbox{NCD}}$, $d^{}$, $d^{\scriptsize \mbox{CDM}}$, {\it CVTree}, $d_{2}^{S}$ and {\it co-phylog} to NGS short reads which were simulated from the whole genome sequences.
Again, even at the very low $1\times$ sampling depth, the clustering results obtained from the NGS short reads were quite similar to those obtained from the whole genome sequences, although both were slightly worse than those obtained from the 16S rRNA sequences (Table 5).
As this experiment was conducted at a high taxonomy level and the species were selected from different orders of the class {\it Gammaproteobacteria}, one could expect that the 16S rRNA sequences should be more suitable for the classification.
It can also be seen from Table 5 that the four distances $d^{\scriptsize \mbox{NCD}}$, $d^{}$, $d^{\scriptsize \mbox{CDM}}$, and {\it CVTree} produced comparably accurate results, and outperformed the other two distances, $d_{2}^{S}$ and {\it co-phylog}, for all three types of sequence data, 16S rRNA sequences, whole genome sequences, and NGS short reads.

In this section we have shown the application of the compression-based distance measures $d^{\scriptsize \mbox{NCD}}$, $d^{}$ and $d^{\scriptsize \mbox{CDM}}$ to classify the genomes of the class {\it Gammaproteobacteria} into their correct orders.
They produced comparably accurate results as the benchmark MSA method and performed consistently well on three types of sequence data.
The $k$-mer based distance {\it CVTree} was also a good choice, but one needs to carefully test a large number of $k$ to find the optimal value for each individual data set.
Even for the same species, the optimal $k$ may not be the same for different types of sequence data, as we have seen for the cases of 16S rRNA sequences and whole genome sequences.
In addition, the optimal $k$ was selected to optimize the parsimony score of the clustering trees.
This will not be possible if we have no prior knowledge about the true classification, which is usually the case in real applications.
The $d_{2}^{S}$ distance showed reasonable performance on 16S rRNA sequences.
However, it failed for the whole genome sequences and NGS short reads, even for optimal $k$.
The {\it co-phylog} distance is not suitable for species with far evolutionary distances from each other.
%

%
%
%
%

\subsection*{Classification of metagenomic samples from mammalian gut reveals the diet and gut physiology of the host species.}

So far we have seen the applications of the alignment-free distance measures for the comparison of genomic sequence data.
Recent studies have suggested that these measures can also be applied for the comparison of metagenomic data, especially with NGS short reads.
%
%
In this section we consider a metagenomic data set that includes NGS short reads of 39 fecal samples from 33 mammalian host species.
The host species can be classified into four groups according to their diet and gut physiology: foregut-fermenting herbivores (13 samples), hindgut-fermenting herbivores (8 samples), carnivores (7 samples), and omnivores (11 samples) (Table S6).
This data set has been studied previously in \cite{muegge_11,jiang_12}.
In \cite{jiang_12} the authors applied the {\it CVTree} and $d_{2}^{S}$ distances to those 39 metagenomic samples and found that the sequence signatures (that is, the $k$-mers) of the samples were strongly associated with the diet and gut physiology of the host species.
Hence we want to test if the compression-based distance measures $d^{\scriptsize \mbox{NCD}}$, $d^{}$ and $d^{\scriptsize \mbox{CDM}}$ can also reveal any interesting results from this metagenomic data set.
%

%
%
%
%
%
%

\subsubsection*{Performance on the sub-data set with 11 omnivore samples excluded}

Following \cite{muegge_11,jiang_12}, we first consider the case when the 11 omnivore samples were excluded due to their complicated microbial compositions.
The remaining 28 samples belong to three groups: foregut-fermenting herbivores, hindgut-fermenting herbivores, and carnivores.
As there is no benchmark tree for this clustering problem, we only used the parsimony score to evaluate the clustering trees.
The optimal parsimony score is 2 and any score lower than 7 corresponds to a $p$-value less than 0.001 (10,000 random trees were generated).
Table 6 shows that the parsimony scores of the {\it CVTree} and $d_{2}^{S}$ distances for optimal $k$ were better than those of the compression-based distances $d^{\scriptsize \mbox{NCD}}$, $d^{}$, and $d^{\scriptsize \mbox{CDM}}$.
We also noted that the parsimony score of the {\it CVTree} distance varied considerably (up to 11), while that of the $d_{2}^{S}$ distance was more stable (Table S7).
%


The optimal tree obtained from the $d_{2}^{S}$ distance ($k=5$, parsimony score = 3) is shown in Figure S6.
Only two samples Rock Hyrax 1 and 2 were wrongly classified to the group of hindgut-fermenting herbivores.
Although the optimal tree obtained from the {\it CVTree} distance ($k=4$) also has the same parsimony score of 3, it has a serious mistake when classifying the two carnivores Polar Bear and Lion to the groups of herbivores (Figure S7).
The clustering tree obtained from the $d^{\scriptsize \mbox{CDM}}$ distance (parsimony score = 5) is shown in Figure S8.
It correctly distinguished carnivores from herbivores.
However, it wrongly classified Rock Hyraxes, Colobus and Visayam Warty Pig to the group of hindgut-fermenting herbivores.

\subsubsection*{Performance on the full data set}

Next, we added back the 11 omnivore samples and repeated the experiment with the full data set.
As there are four groups in the true classification, the optimal parsimony score is 3 and any score lower than 15 corresponds to a $p$-value less than 0.001 (10,000 random trees were generated).
We found that the best parsimony score was obtained from the $d^{\scriptsize \mbox{NCD}}$ distance, followed by {\it CVTree} ($k=6$) and $d_{2}^{S}$ ($k=7$) (Table 6).
It should be noted that the optimal $k$ of the {\it CVTree} and $d_{2}^{S}$ distances for the sub-data set and for the full data set were different (Table 6, Table S7).
The clustering tree of the $d^{\scriptsize \mbox{NCD}}$ distance is shown in Figure 4.
The samples from foregut-fermenting herbivores  were well clustered together, except for Rock Hyraxes, Colobus, and Visayam Warty Pig, which were classified to the group of hindgut-fermenting herbivores.
This is similar to the earlier observation when the 11 omnivore samples were excluded.
Figure 4 also shows that the carnivore samples were grouped together.
The omnivore samples, however, were scattered throughout the groups of herbivores and carnivores.
This again indicates the diversity of the gut microbial communities of omnivores, as mentioned previously in \cite{muegge_11,jiang_12}.
Another important observation from Figure 4 is that the samples from primates, including Baboon 1 and 2, Chimpanzee 1 and 2, Orangutan, Gorilla, Callimicos, Saki, Black Lemur, were clustered together into one group.
This may suggest that those primates share common features in their gut microbial environments.
Finally, it can also be seen that two samples of the same host species were often clustered close to each other such as Chimpanzee 1 and 2, Lion 1 and 2, Okapi 1 and 2, Bighorn Sheep 1 and 2, supporting the accuracy of the classification and the $d^{\scriptsize \mbox{NCD}}$ distance.

The results obtained in this section have demonstrated another application of the alignment-free measures of sequence distance: comparison and classification of metagenomic samples of NGS short reads.
This task is of critical importance for the understanding of microbial communities.
Both $k$-mer based and compression-based distance measures have revealed interesting results about the microbial communities of mammalian gut from their metagenomic samples.
In particular, the information contained in the samples was strongly associated with the diet and gut physiology of herbivores, carnivores, and omnivores.
This agrees well with previous studies in \cite{muegge_11,jiang_12}.
Moreover, our results obtained from the compression-based distance measures also discovered a strong similarity between the gut microbial communities of the primates.
This interesting finding has not been observed in previous studies.
%
%

\section*{Conclusions}

In this paper we studied the application of the compression-based distance measures for the problem of sequence comparison with a special focus on NGS short reads data.
The key advantages of the compression-based distance measures are assembly-free, alignment-free, and parameter-free.
We conducted extensive validation on various types of sequence data: 16S rRNA sequences, mtDNA sequences, whole genome sequences, and NGS short reads
The sequence data came from several mammalian and bacteria genomes at different taxonomy levels, as well as from microbial metagenomic samples.
The results show that the compression-based distance measures produced comparably accurate results as the $k$-mer based methods, and both had good agreement with the alignment-based approach and existing benchmarks in the literature.

The $k$-mer based distance measures, however, may produce inconsistent results depending on the parameter $k$, the type of sequence data, or the species under consideration.
For example, the {\it co-phylog} measure was not applicable to species with far evolutionary distances from each other (data set of 29 mammalian, data set of 70 {\it Gammaproteobacteria}, data set of 39 metagenomic samples).
The $d_{2}^{S}$ measure failed for the data set of 29 {\it Escherichia/Shigella} bacteria, for whole genome sequences and NGS short reads of the data set of 70 {\it Gammaproteobacteria}.
The {\it CVTree} measure produced inconsistent results for the data set of 29 {\it Escherichia/Shigella} bacteria.
The compression-based measures, although did not always provide the best distances, but  performed consistently well across all data sets in the study.
This is the result of the parameter-free feature of the compression-based measures.
On the other hand, choosing the optimal parameter $k$ for each data set is of critical importance for using the $k$-mer based methods.
This task may be a difficult problem when there is no benchmark (e.g., true phylogenetic trees, true classifications) available to guide the analysis and the selection of $k$.

One possible drawback of the compression-based distance measures is the running time.
Obviously, compressing a DNA sequence (or an NGS sample) takes longer time than counting its $k$-mers.
Moreover, the compression-based methods need to perform pairwise compression of the input sequences, whereas the $k$-mer methods only need to calculate one frequency vector for each input sequence.
However, it should be also noted that in general one may need to test a wide range of $k$ to find the optimal results when using the $k$-mer methods.
%
%
%
For instance, for the data set of 39 metagenomic samples in our study, the running time of the {\it CVTree} measure was $\sim$1-7 minutes for each $k=2, 3, \ldots, 10$,  and $\sim$10-60 minutes for each $k=11, 12, \ldots, 20$, etc.
Thus, a test covering all values of $k=2, 3, \ldots, 10$ only took less than 20 minutes, but to include other values of $k=11, 12, \ldots, 20$, the running time was dramatically increased up to $\sim$8 hours.
%
%
The running time of the compression-based measures for this data set was $\sim$25 hours, about 3 times longer than that of {\it CVTree}.
Faster compression tools, especially those developed specifically for NGS short reads such as {\it BEETL} \cite{cox_12}, {\it SCALCE} \cite{hach_12}, etc, can be applied to improve the running time.
However, one should be also aware of the trade-off between the compression time  and the compression efficiency.
More efficient compression yields more accurate distance but takes longer running time.
Our future research will focus on improving the running time and studying the effects of different compression tools.

To the best of our knowledge, our work is the first study to assess the performance of the compression-based distance measures on NGS short reads data.
More importantly, our results have demonstrated the accuracy and the consistency of these measures on both NGS short reads and long genome sequences.
These findings underscore the advantages of the compression-based distance measures, suggesting that these measures also represent powerful tools for the alignment-free sequence comparison, in addition to the popular $k$-mer methods.

\section*{Acknowledgments}

\section*{}
\bibliography{plos1_article}

\clearpage
\section*{Figure Legends}

\begin{figure}[!ht]
\begin{center}
\end{center}
\caption{
{\bf Phylogenetic trees reconstructed from NGS short reads of 29 mtDNA sequences using: (a) $d^{\scriptsize \mbox{CDM}}$, (b) {\it CVTree} $(k=10)$, (d) $d_{2}^{S} (k=8)$.}
The short reads were simulated from the tool {\it MetaSim} using the Empirical model and $5\times$ sampling depth. The group of three species platypus, opossum, and wallaroo was used as the outgroup to root the tree.
}
\label{Figure1}
\end{figure}

\begin{figure}[!ht]
\begin{center}
\end{center}
\caption{
{\bf Phylogenetic trees reconstructed from 29 {\it Escherichia/Shigella} genomes using (a) {\it co-phylog}, (b) $d^{\scriptsize \mbox{CDM}}$, and from NGS short reads using (c) $d^{\scriptsize \mbox{CDM}}$.}
The short reads were simulated from the tool {\it MetaSim} using the Exact model and $1\times$ sampling depth. {\it Escherichia Fergusonii} was used as the outgroup to root the tree.
}
\label{Figure2}
\end{figure}

\begin{figure}[!ht]
\begin{center}
\end{center}
\caption{
{\bf Clustering tree reconstructed from 16S rRNA sequences of 70 {\it Gammaproteobacteria} genomes using the $d^{\scriptsize \mbox{NCD}}$ distance.}
Those 70 genomes belong to 15 orders which are indicated by different colors in the figure.
}
\label{Figure3}
\end{figure}

\begin{figure}[!ht]
\begin{center}
\end{center}
\caption{
{\bf Clustering tree reconstructed from 39 metagenomic samples using the $d^{\scriptsize \mbox{NCD}}$ distance.}
The host species' colors indicate their diet and gut physiology: foregut-fermenting herbivores (green), hindgut-fermenting herbivores (yellow), carnivores (red) and omnivores (blue).
}
\label{Figure4}
\end{figure}

\clearpage
\section*{Tables}

\begin{table}[!ht]
\caption{
{\bf Comparison of alignment-free distances and MSA distance for NGS short reads of mtDNA sequences.}}
\begin{tabular}{c c c c c c}
  \hline
  & $d^{\scriptsize \mbox{NCD}}$ & $d^{}$ & $d^{\scriptsize \mbox{CDM}}$ & {\it CVTree} ($k=10$) & $d_{2}^{S} (k=8)$ \\ \hline
  454 & 14 & 16 & {\bf 10} & {\bf 10} & {\bf 6} \\
  Exact & {\bf 8} & {\bf 8} & {\bf 8} & {\bf 8} & {\bf 8} \\
  Empirical & {\bf 6} & 8 & {\bf 4} & 8 & 12 \\
  Sanger & {\bf 10} & 14 & {\bf 8} & 14 & {\bf 8} \\
  \hline
  454 & 0.68 & 0.66 & 0.68 & {\bf 0.75} & {\bf 0.88} \\
  Exact & 0.69 & 0.68 & 0.68 & {\bf 0.71} & {\bf 0.88} \\
  Empirical & {\bf 0.69} & {\bf 0.69} & 0.66 & {\bf 0.69} & {\bf 0.81} \\
  Sanger & 0.67 & 0.67 & 0.65 & {\bf 0.74} & {\bf 0.87} \\
  \hline
\end{tabular}
\begin{flushleft}
The short reads were simulated from the mtDNA sequences using four error models 454, Exact, Empirical, and Sanger of the tool {\it MetaSim} at $5\times$ sampling depth. The two smallest tree symmetric differences and the two highest distance correlation coefficients for each error model are highlighted in boldface. Similar results for $1\times$, $10\times$, and $30\times$ sampling depths can be found in Table S2.
\end{flushleft}
\label{tab:Table1}
\end{table}

\begin{table}[!ht]
\caption{
{\bf Comparison of phylogenetic trees reconstructed from mtDNA sequences and from NGS short reads.}}
\begin{tabular}{c c c c c c}
  \hline
  & $d^{\scriptsize \mbox{NCD}}$ & $d^{}$ & $d^{\scriptsize \mbox{CDM}}$ & {\it CVTree} ($k=10$) & $d_{2}^{S} (k=8)$ \\ \hline
  454 & 10 & 14 & {\bf 8} & {\bf 8} & {\bf 4} \\
  Exact & {\bf 2} & {\bf 0} & 4 & {\bf 2} & {\bf 2} \\
  Empirical & {\bf 8} & {\bf 6} & {\bf 6} & {\bf 6} & 10 \\
  Sanger & 12 & {\bf 10} & {\bf 8} & {\bf 10} & {\bf 8} \\
  \hline
\end{tabular}
\begin{flushleft}
The short reads were simulated from the mtDNA sequences using four error models 454, Exact, Empirical, and Sanger of the tool {\it MetaSim} at $5\times$ sampling depth. The two smallest tree symmetric differences for each error model are highlighted in boldface. Similar results for $1\times$, $10\times$, and $30\times$ sampling depths can be found in Table S2.
\end{flushleft}
\label{tab:Table2}
\end{table}

\begin{table}[!ht]
\caption{
{\bf Comparison of alignment-free distances and the benchmark {\it co-phylog} distance for 29 {\it Escherichia/Shigella} genomes.}}
\begin{tabular}{c c c c c c c}
  \hline
  & $d^{\scriptsize \mbox{NCD}}$ & $d^{}$ & $d^{\scriptsize \mbox{CDM}}$ & {\it CVTree} ($k=9$) & {\it CVTree} ($k=21$) & $d_{2}^{S} (k=8)$ \\
  \hline
  tree symmetric difference & 16 & {\bf 14} & {\bf 12} & 20 & 16 & 24 \\
  distance correlation & {\bf 0.97} & 0.95 & {\bf 0.99} & 0.80 & 0.80 & 0.20 \\
  \hline
\end{tabular}
\begin{flushleft}
The two smallest tree symmetric differences and the two highest correlation coefficients are highlighted in boldface.
\end{flushleft}
\label{tab:Table3}
\end{table}

\begin{table}[!ht]
\caption{
{\bf Comparison of alignment-free distances and the benchmark {\it co-phylog} distance for NGS short reads of 29 {\it Escherichia/Shigella} genomes.}}
\begin{tabular}{c c c c c c c c}
  \hline
  & $d^{\scriptsize \mbox{NCD}}$ & $d^{}$ & $d^{\scriptsize \mbox{CDM}}$ & {\it CVTree} ($k=9$) & {\it CVTree} ($k=15$) & {\it CVTree} ($k=21$) & $d_{2}^{S} (k=8)$ \\ \hline
  454 & 16 & {\bf 14} & {\bf 12} & 22 & 22 & 54 & 50 \\
  Exact & {\bf 12} & {\bf 10} & {\bf 10} & 22 & 24 & 36 & 50 \\
  Empirical & 16 & {\bf 12} & {\bf 14} & 24 & 18 & 42 & 54 \\
  Sanger & {\bf 12} & {\bf 10} & {\bf 10} & 28 & 24 & 40 & 50 \\
  \hline
  454 & {\bf 0.96} & {\bf 0.96} & {\bf 0.97} & 0.74 & 0.87 & 0.31 & 0.04 \\
  Exact & {\bf 0.97} & {\bf 0.97} & {\bf 0.98} & 0.82 & 0.82 & 0.61 & 0.05 \\
  Empirical & {\bf 0.96} & {\bf 0.96} & {\bf 0.96} & 0.77 & 0.85 & 0.38 & 0.01 \\
  Sanger & {\bf 0.97} & {\bf 0.97} & {\bf 0.97} & 0.78 & 0.86 & 0.36 & 0.04 \\
  \hline
\end{tabular}
\begin{flushleft}
The short reads were simulated from the {\it Escherichia/Shigella} genomes using four error models 454, Exact, Empirical, and Sanger of the tool {\it MetaSim} at $1\times$ sampling depth. The two smallest tree symmetric differences and the two highest correlation coefficients for each error model are highlighted in boldface. Similar results for $5\times$ sampling depth can be found in Table S3.
\end{flushleft}
\label{tab:Table4}
\end{table}

\begin{table}[!ht]
\caption{
{\bf Comparison of alignment-free distances and the benchmark MSA distance for 70 {\it Gammaproteobacteria} genomes.}}
\begin{tabular}{c c c c c c c c}
  \hline
  & & $d^{\scriptsize \mbox{NCD}}$ & $d^{}$ & $d^{\scriptsize \mbox{CDM}}$ & {\it CVTree} & $d_{2}^{S}$ & {\it co-phylog} \\
  \hline
  & parsimony score & {\bf 17} & {\bf 18} & {\bf 18} & {\bf 17}& {\bf 18} & 25\\
  16s rRNA sequences & tree symmetric difference & {\bf 50} & {\bf 52} & {\bf 52} & {\bf 50} & 62 & 108\\
  & distance correlation & {\bf 0.93} & 0.90 & {\bf 0.93} & {\bf 0.92} & {\bf 0.92} & 0.65\\
  \hline
  & parsimony score & {\bf 22} & {\bf 22} & {\bf 21} & {\bf 21} & 31 & 26\\
  genome sequences & tree symmetric difference & 80 & {\bf 78} & {\bf 76} & 84 & 110 & 110\\
  & distance correlation & 0.47 & 0.46 & 0.47 & {\bf 0.67} & {\bf 0.50} & 0.45\\
  \hline
  & parsimony score & {\bf 21} & {\bf 19} & 23 & 24 & 32 & 28\\
  NGS short reads & tree symmetric difference & 90 & {\bf 70} & {\bf 84} & 88 & 114 & 116\\
  & distance correlation & {\bf 0.60} & 0.58 & 0.53 & {\bf 0.63} & 0.48 & 0.42\\
  \hline
\end{tabular}
\begin{flushleft}
The NGS short reads were simulated from the whole genome sequences using the Exact model of {\it MetaSim} at $1\times$ sampling depth . The two smallest parsimony scores, the two smallest tree symmetric differences and the two highest correlation coefficients are highlighted in boldface. For {\it CVTree}, we used $k=7$ for the 16S rRNA data set and $k=12$ for the whole genome and NGS data sets. For $d_{2}^{S}$, we used $k=6$ for the 16S rRNA data set and $k=8$ for the whole genome and NGS data sets.
\end{flushleft}
\label{tab:Table5}
\end{table}

\begin{table}[!ht]
\caption{
{\bf Parsimony score for the classification of 39 metagenomic samples using the alignment-free distances.}}
\begin{tabular}{l c c c c c}
  \hline
  & $d^{\scriptsize \mbox{NCD}}$ & $d^{}$ & $d^{\scriptsize \mbox{CDM}}$ & {\it CVTree} & $d_{2}^{S}$ \\
  \hline
  sub-data set (omnivore samples excluded) & 6 & 7 & {\bf 5} & {\bf 3} & {\bf 3} \\
  full data set & {\bf 9} & 12 & 12 & {\bf 10} & {\bf 10} \\
  \hline
\end{tabular}
\begin{flushleft}
For {\it CVTree}, we used $k=6$ for the full data set and $k=4$ for the sub-data set in which the omnivore samples were excluded. For $d_{2}^{S}$, we used $k=7$ for the full data set and $k=5$ for the sub-data set. The two smallest parsimony scores for each data set are highlighted in boldface.
\end{flushleft}
\label{tab:Table6}
\end{table}

\clearpage
\section*{Supplementary Figure Legends}

\noindent{\bf Figure S1.} Phylogenetic trees reconstructed from 29 mtDNA sequences using: (a) MSA, (b) $d^{\scriptsize \mbox{CDM}}$, (c) {\it CVTree} $(k=10)$, (d) $d_{2}^{S} (k=8)$.

\noindent{\bf Figure S2.} Phylogenetic trees reconstructed from 29 {\it Escherichia/Shigella} genomes using: (a) {\it co-phylog}, (b) {\it CVTree} $(k=9)$, (c) {\it CVTree} $(k=15)$, (d) {\it CVTree} $(k=21)$.

\noindent{\bf Figure S3.} Clustering tree reconstructed from 16S rRNA sequences of 70 {\it Gammaproteobacteria} genomes using the MSA distance.

\noindent{\bf Figure S4.} Clustering tree reconstructed from 16S rRNA sequences of 70 {\it Gammaproteobacteria} genomes using the distance {\it CVTree} ($k=7$).

\noindent{\bf Figure S5.} Clustering tree reconstructed from 16S rRNA sequences of 70 {\it Gammaproteobacteria} genomes using the distance $d_{2}^{S}$ ($k=6$).

\noindent{\bf Figure S6.} Clustering tree reconstructed from the metagenomic samples (omnivore samples excluded) using the distance $d_{2}^{S} (k=5)$.

\noindent{\bf Figure S7.} Clustering tree reconstructed from the metagenomic samples (omnivore samples excluded) using the distance {\it CVTree} $(k=4)$.

\noindent{\bf Figure S8.} Clustering tree reconstructed from the metagenomic samples (omnivore samples excluded) using the distance $d^{\scriptsize \mbox{CDM}}$.

\clearpage
\section*{Supplementary Table Legends}

\noindent{\bf Table S1.}  Comparison of alignment-free distances and the benchmark MSA distance for 29 mtDNA sequences.

\noindent{\bf Table S2.}  Comparison of alignment-free distances and the benchmark MSA distance for 29 mtDNA sequences and their short reads.

\noindent{\bf Table S3.}  Comparison of alignment-free distances and the benchmark {\it co-phylog} distance for 29 {\it Escherichia/Shigella} genomes and their short reads.

\noindent{\bf Table S4.}  The list of 70 genomes in the class {\it Gammaproteobacteria}, their orders, and their accession numbers.

\noindent{\bf Table S5.}  Comparison of alignment-free distances and the benchmark MSA distance for 70 {\it Gammaproteobacteria} genomes and their short reads.

\noindent{\bf Table S6.}  The list of 39 metagenomic samples and their host species' diet and gut physiology.

\noindent{\bf Table S5.}  The parsimony score of the clustering trees for 39 metagenomic samples.

\end{document}